\documentclass[aps,prl,onecolumn,superscriptaddress,floatfix,]{revtex4-2}

\usepackage{amsmath,amssymb}
\usepackage{graphicx}
\usepackage{dcolumn}
\usepackage{bm}
\usepackage{hyperref}
\usepackage{xcolor}
\usepackage{physics}
\usepackage{float}

\begin{document}

\title{On-Demand Coherent Mapping of Telecom Optical States onto Erbium Hyperfine Spins}

\author{Zongfeng Li}
\affiliation{Department of Electrical and Computer Engineering, Northwestern University, Evanston, IL U.S.A}

\author{Mahdi Hosseini}
\thanks{Email: mh@northwestern.edu}
\affiliation{Department of Electrical and Computer Engineering, Northwestern University, Evanston, IL U.S.A} \affiliation{Applied Physics Program, Northwestern University,  Evanston, IL U.S.A}

\date{\today}
\begin{abstract}

Optical quantum memories operating directly at telecom wavelengths are a key enabling technology for long-distance quantum networks, yet on-demand storage onto long-lived ground-state spins in this spectral region has remained elusive due to the challenge of coherently transferring optical excitations to  hyperfine spin states. Here we demonstrate spin-wave storage in $^{167}$Er$^{3+}$:Y$_2$SiO$_5$ at 0.8~K and 1.1~T, establishing the core operational primitive required for on-demand telecom quantum memories. Using classical optical control pulses, we coherently transfer collective optical excitations to erbium hyperfine states with transfer efficiency exceeding 12\%, enabling on-demand retrieval. We measure a hyperfine population lifetime of 25~s and demonstrate spin-wave storage for up to 25~$\mu$s. By identifying hyperfine inhomogeneous broadening as the dominant present limitation, our measurements define a clear pathway toward second-scale storage through improved spectral tailoring and dynamical decoupling. The results highlight the application of erbium-based solid-state memories for scalable fiber-compatible quantum repeater architectures.

\end{abstract}

\maketitle

\section*{Introduction}
%

Future quantum network infrastructure require efficient,
low-noise optical quantum memories~\cite{Lei2023} that are compatible with the telecom C-band used in deployed fiber networks, enabling synchronization of probabilistic photonic entanglement generation without lossy frequency conversion interfaces. 

Rare-earth-ion-doped crystals are among the most promising platforms for optical quantum memories. Rare-earth solids' large inhomogeneous broadening enables intrinsically multimode and bandwidth-multiplexed storage\cite{Sinclair:2014aa, Lei:2025aa, Li:2025aa, wei2024quantum, Dajczgewand:2015aa}, and their non-zero nuclear spin enables the prospect of hyperfine coherence times reaching seconds \cite{Rancic2018, ma2021one, wang2025nuclear, OSullivan:2025aa}. Among rare-earth ions, $^{167}$Er$^{3+}$ is particularly interesting because it combines optical transitions in the low-loss telecom C-band with long-lived hyperfine spin states.  Erbium-based memories are being considered as a compelling platform for scalable fiber-compatible quantum repeater nodes and integrated cryogenic quantum photonic architectures\cite{craiciu2021multifunctional, Yang:2023aa}. Although cryogenic operation is often cited as a limitation for scalability, these solid-state memories often operate inside a table-top cryostat ($>$0.5K) co-hosting superconducting nanowire single-photon detectors, and support miniaturization and scalable nanofabrication\cite{Dutta:2023aa, Yang:2023aa}.

In typical quantum storage using rare-earth solids~\cite{Afzelius2009}, the storage time is fixed precluding on-demand readout, while the storage time is also limited by the optical coherence lifetime. Extension to spin-wave storage overcomes these limitations by transferring the excitation to long-lived spin states, although in practice the coherence is often constrained by electron-spin dephasing unless nuclear-spin degrees of freedom are accessed.
Despite this promise, realization of on-demand telecom storage in erbium has not been achieved to date, because it requires coherent transfer of collective optical excitations into long-lived hyperfine spin states with high phase fidelity. 

On-demand and long-lived storage requires the optical coherence to be mapped into a
spin coherence (spin wave) via $\pi$ pulses, analogous to other three-level memory systems \cite{Duan2001, Hsiao2018, Hosseini:2009aa}.
Along these lines, spin-wave atomic-frequency comb (AFC) storage\cite{Afzelius2009} has been demonstrated in Pr:YSO~\cite{Jobez2015},
Eu:YSO~\cite{Laplane2017, zhu2024integrated}, and Tm:LiNbO$_3$~\cite{Saglamyurek2011}.
To date, a variety of protocols has been demonstrated in Erbium doped crystals \cite{ranvcic2018coherence, wei2024quantum, liu2022demand, craiciu2021multifunctional, dibos2018atomic}, but the challenge of spin-wave storage still remains unresolved.
The reasons why it's particularly hard in Er$^{3+}$ include: 1) a much lower temperature and stronger magnetic field is required to remove the phonon coupling due to its unpaired electron (Kramers ion) \cite{ranvcic2018coherence}; 2) a large number of hyperfine state induce population dilution and transition complexity \cite{stuart2021initialization}; 3) Er$^{3+}$ exhibits weak hyperfine state mixing in a strong field, resulting in a reduced branching ratio and consequently a less efficient $\Lambda$ system. This behavior arises because the Zeeman splitting significantly exceeds the off-diagonal mixing terms such as $\mathbf{I} \cdot \mathbf{Q} \cdot \mathbf{I}$ and $\mathbf{S} \cdot \mathbf{A} \cdot \mathbf{I}$ in the Hamiltonian. 

Beyond on-demand retrieval, operating in the spin-wave regime unlocks storage times limited by the hyperfine $T_2$ rather than the optical $T_2$.
Recent measurements in $^{167}$Er:YSO demonstrated hyperfine coherence
times exceeding 1~s when dynamical decoupling is
applied~\cite{Rancic2018}, far exceeding the sub-millisecond optical $T_2$.
Realizing spin-wave storage in Er:YSO therefore represents a critical
milestone toward a telecom quantum memory with both on-demand retrieval
and second-scale storage, two features that must coexist for practical
quantum communication nodes.

Here, we report the realization of spin-wave storage in Er:YSO. By identifying effective $\Lambda$-system, realizing efficient optical pumping, and using strong optical control pulses, we achieve coherent population transfer between ground-state hyperfine levels. This is a demonstration of the missing control primitive required for on-demand quantum storage directly in the telecom band. We investigate the dependence of the retrieval efficiency on the properties of the control pulses and characterize the coherence time of the hyperfine spin states. Our results define the engineering requirements for long-lived multiplexed telecom quantum memories and suggest a pathway toward quantum repeater operation through spectral tailoring and dynamical decoupling.

\begin{figure*}
    \centering
    \includegraphics[width=0.8\linewidth]{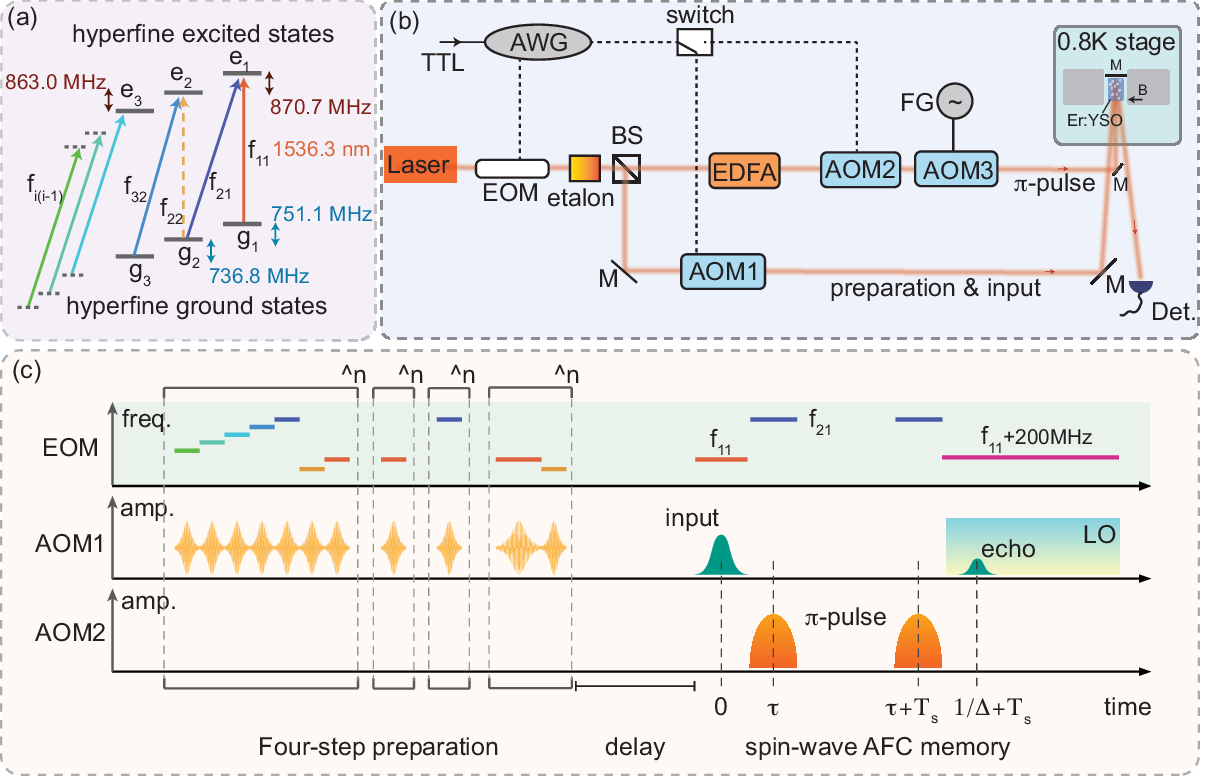}
    \caption{%
    (a)~Schematic of the hyperfine energy levels of Er$^{3+}$:YSO (site 1) at B = 1.1 T showing
    ground ($g_i$) and excited ($e_i$) manifolds.
    The AFC and $\pi$-pulse is applied at transition $g_1 \leftrightarrow e_1$ and $g_2 \leftrightarrow e_1$, respectively.
    (b) and (c)~Schematic of the experimental setup and the pulse sequence.
    The probe and the control-pulse optical beams enter the crystal in
    a cryostat at a relative angle
    $\theta \approx $~20~mrad to allow spatial filtering of the echo. An Electro-optic modulator (EOM) and an etalon filter are used for tuning and filtering the laser frequency to address $f_{ij}$ transitions.
    The control ($\pi$) pulses are generated by amplifying a fraction of the probe laser
    in an Er-doped fiber amplifier (EDFA), followed by acousto-optical modulators (AOMs) to create pulses and suppress leakage.
    The four-step preparation creates an AFC in $g_1$ and empty $g_2$, followed by the spin-wave AFC storage via the $g_1$-$e_1$-$g_2$ $\Lambda$-system.
    The echo is detected by heterodyne detection with a lock-in amplifier where the local oscillator (with frequency difference of 200~MHz) co-propagates with the probe. Delay times up to 100~ms were applied after preparation.
  }
  \label{fig:level_setup}
\end{figure*}
\section{Platform and Results}

Isotopically purified $^{167}$Er$^{3+}$ doped in YSO at non-zero magnetic field exhibits a rich hyperfine manifold.
For $^{167}$Er, the $^4I_{15/2}$ ground
state and $^4I_{13/2}$ excited state each split into eight hyperfine levels
($g_1-g_8$ and $e_1-e_8$, respectively) at the applied field of about 1.1 T (Fig.~\ref{fig:level_setup}(a)).

The experimental setup and the experimental pulse sequence are shown in Fig.~\ref{fig:level_setup}(b)-(c).
The probe and pump pulses (input pulse and AFC preparation pulses) and the control pulses ($\pi$ pulses) 
enter the crystal at a small angle. After the crystal, spatial filtering in the far field separates the probe from the strong control pulses. Additional details about the crystal and operating conditions can be found in our previous work,\cite{Li:2025aa}. 

The control pulses are generated by routing a fraction of the probe laser through an Er-doped fiber amplifier (EDFA), with acousto-optic modulators (AOMs) placed after the amplifier to control the pulse amplitude and suppress the leakage. The photon echo is measured directly by photodiode or via heterodyne detection using a local oscillator (LO) detuned by 200 MHz (generated by EOM) and co-propagating with the probe. The resulting beat signal is demodulated with a lock-in amplifier (see Methods), enabling shot-noise-limited sensitivity at the echo frequency. Because the AFC protocol is intrinsically linear, preservation of coherence in the bright-pulse regime implies compatibility with weak coherent and single-photon storage, establishing the essential operational requirement for quantum-memory functionality.
The AOM shapes the amplitude of the input and creates the secant hyperbolic pulses \cite{jobez2016towards} for preparation and control pulses. The EOM tunes the laser to different transition frequencies.

\begin{figure}
    \centering    \includegraphics[width=0.8\linewidth]{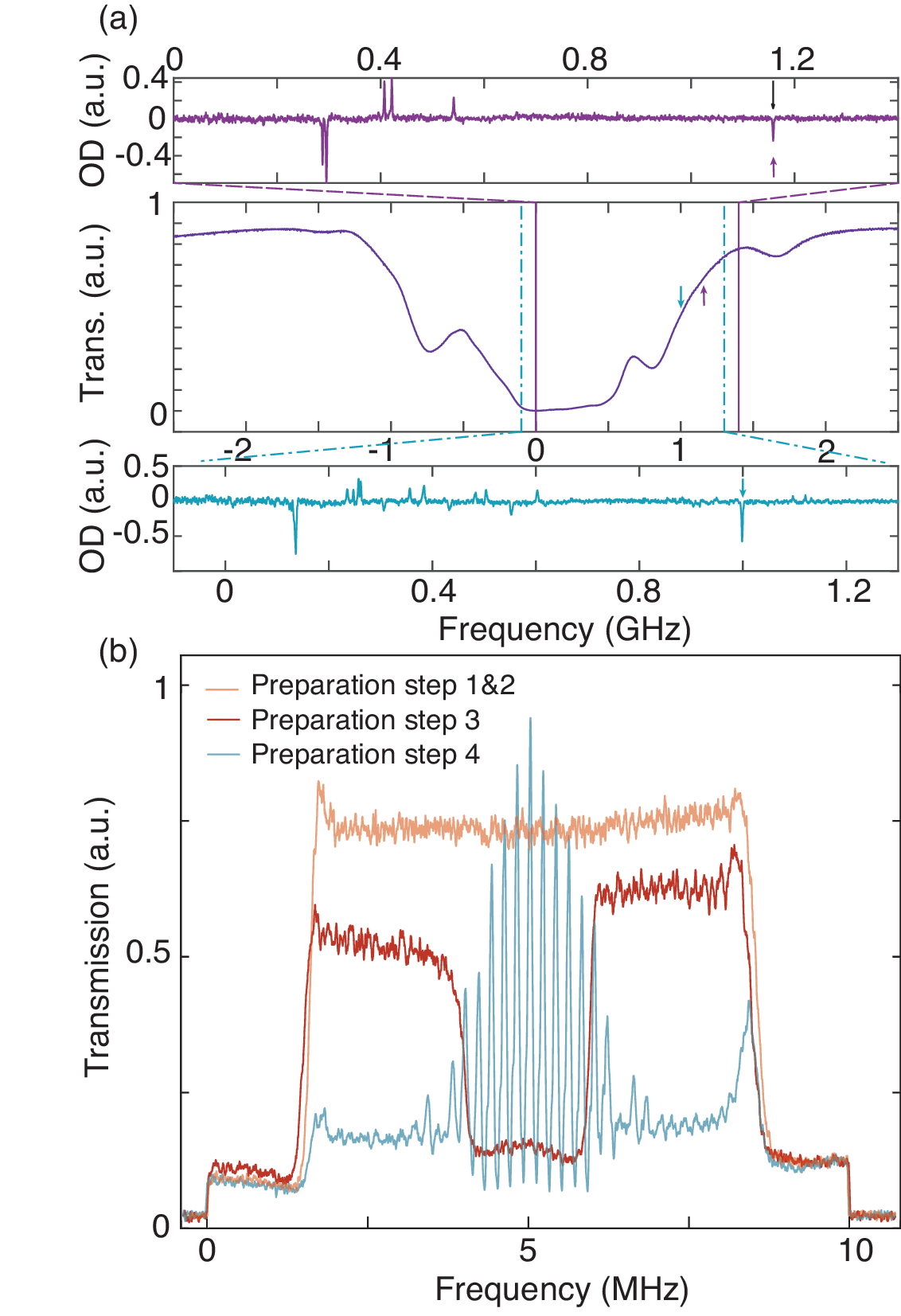}
    \caption{
    (a) In the hole-burning curves (top and bottom), antihole(peak) structures appear when pumping at the right edge of the $^4I_{15/2} \rightarrow ^4I_{15/2}$ absorption line (middle). The antiholes' frequencies determine the $\Lambda$ system and the hyperfine structure.
    (b) Result of the four-step preparation used for effective initialization of an atomic frequency comb (AFC) into $g_1$ state.
    Steps 1 and 2: a 7-MHz-wide class cleaning and initialization in $g_2$ state. Step 3: 2-MHz-wide population is burned from $g_2$ to $g_1$. Step 4: a 2-MHz-AFC is burned at $g_1$ state meanwhile cleaning $g_2$ level. The $g_1$ population is characterized by the transition $f_{11}$.
    }
  \label{fig2}
\end{figure}

To identify the $\Lambda$ system required for spin-wave storage, we performed the hole-burning using transitions  $g_i \rightarrow $$e_{i-1}$ (denote as $f_{i(i-1)}$) at the edge of the absorption line.
Figure~\ref{fig2}(a) shows two examples of hole-burning: at the edge of inhomogeneous broadening (upper curve), the smaller number of antiholes indicates that ions with only higher hyperfine levels ($f_{21},f_{32}$) are in resonant with the pump; when closer to the center (lower curve), more hyperfine levels are engaged and more antiholes appear. Those antiholes reveals the hyperfine structures of $^4I_{15/2}$ and $^4I_{13/2}$ (see Supp. Note S1).

Figure~\ref{fig2} (b) compares three preparation procedures (see Methods) that achieve efficient AFC in level $g_1$. 
The ground-state hyperfine relaxation time $T^{(g)}_1$ is still $\sim$ 3~s \cite{li2025efficient}, although an antihole lifetimes of up to 30~s is observed (see Supp. Note S2).

Although a fraction of the ions relax back to the $f_{11}$ transition during preparation, as evidenced by the lower spectral plateau in Fig.~\ref{fig2}(b), this does not adversely affect spin-wave storage. This is because the ions participating in the desired $\Lambda$ system are repumped at step 4, while ions outside the 
$\Lambda$ system (due to inhomogeneous broadening) do not contribute noise at the spin-wave echo time.


Given the small dipole moment of Er ions, the optical power needed to create effective control pulses is on the order of one Watt (see Methods). Moreover, the hyperfine ground-state levels of Er:YSO are inhomogeneously broadened
owing to local strain, crystal-field disorder, and superhyperfine coupling
to neighboring nuclei.
This ground-state hyperfine inhomogeneous linewidth $\Gamma_{\rm inh}^{(g)}$ is
distinct from the optically measured $T_2$~\cite{Rancic2018} and is
characterized by the spectral antihole width $\delta\nu_{\rm hole}$ burned
on the $g_1$--$e_1$ transition:
\begin{equation}
  \delta\nu_{\rm hole} \;\approx\; \frac{1}{\pi T_2} + \Gamma_{\rm inh}^{(g)},
  \label{eq:hole_width}
\end{equation}
where the second term accounts for the dephasing of atoms returned to the
ground state~\cite{Macfarlane2002}.
A wide hole width implies a short effective $T_2^* = 1/(\pi \Gamma_{\rm inh}^{(g)})$
for the ground-state hyperfine coherence, which in turn demands a
$\pi$-pulse spectral bandwidth exceeding $\Gamma_{\rm inh}^{(g)}$
to drive all atoms participating in the spin wave.

\begin{figure*}
    \centering
\includegraphics[width=0.85\linewidth]{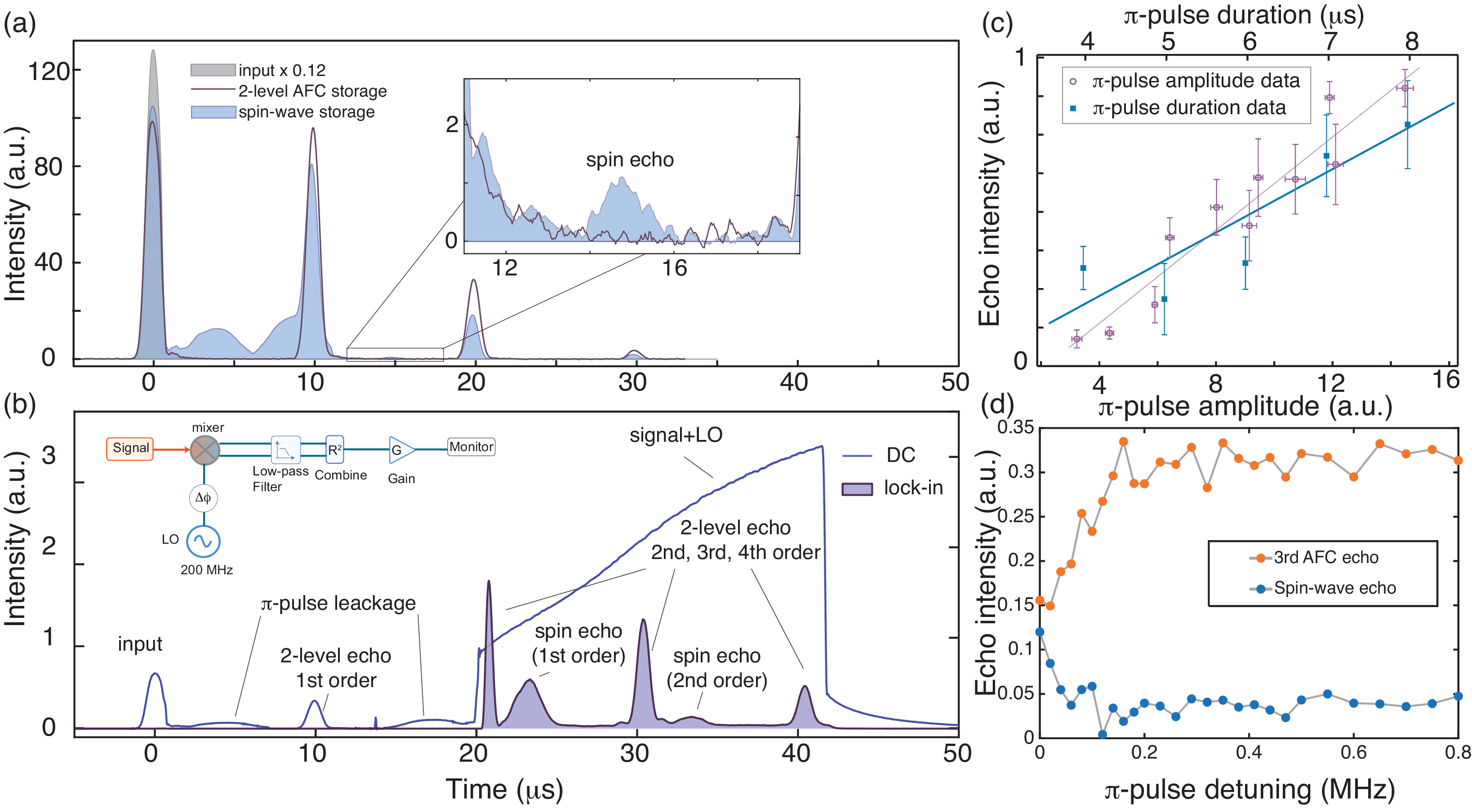}
    \caption{(a) Typical 2-level (without $\pi$-pulses) and spin-wave storage data are shown as measured with a photo-detector (DC), with $T_{AFC}=10\ \mu\mathrm{s},\ T_s=5\ \mu\mathrm{s}$. Inset shows the amplified spin-wave echo. (b) Spin-wave storage data is shown for both DC and lock-in detection, with $T_{AFC}=10\ \mu\mathrm{s},\ T_s=13\ \mu\mathrm{s}$. A local oscillator (200~MHz detunned) co-propagates with the probe pulses during 20-42$\mu$s time frame. Inset shows the lock-in detection setup.(c) Measured echo amplitude as a function of the control $\pi$ pulse duration and strength. Solid lines are linear fits. (d) Intensities of the spin-wave echo and the 3rd AFC 2-level echo measured for different control-pulse detunings from the excited state. }
    \label{fig3}
\end{figure*}

Figure~\ref{fig3}(a) shows a typical 2-level and spin-wave storage measured directly with a photodetector, with total storage time of 15 $\mu$s. 
Figure~\ref{fig3}(b) shows a typical time-domain trace of spin-wave storage measured using lock-in detection.  The storage of an input pulse is carried out in an AFC with tooth spacing $\Delta = $ 0.1 MHz.
In addition to the expected sequence of two-level echoes at
$t = 1/\Delta,\; 2/\Delta,\; 3/\Delta$, a spin-wave echo is clearly
visible at $t = 1/\Delta + T_s$ with total storage time about 23 $\mu$s.
The two-level echo sequence persists because the $\pi$-pulse efficiency
is below unity at the current power level. As efficiency improves, the
two-level echoes are predicted to be suppressed~\cite{Afzelius2009}. We estimate a transfer efficiency of 12\% for a single control pulse, by comparing the 2-level echo and spin-wave echo without lock-in detection.
The 2-level AFC is measured to be 9\%, benefiting from the enhanced absorption by the spin initialization compared to our previous work\cite{li2025efficient}.

Figure~\ref{fig3}(c) plots the spin-wave echo intensity as a function
of $\pi$-pulse amplitude and duration.
The echo intensity increases monotonically with $\pi$-pulse amplitude,
consistent with an increasing spin-transfer efficiency $\eta_\pi$.
Importantly, no saturation or decoherence-induced roll-off is observed
over the explored range, indicating that the $\pi$ pulses do not
introduce significant extra dephasing at these powers.
The linear scaling of the echo intensity with $\pi$-pulse area is
consistent with Eq.~\eqref{eq:pi_bandwidth} in the regime $\Omega_R \tau_\pi < \pi$.\\

Figure~\ref{fig3}(d) shows the intensity of spin-wave echo and 2-level AFC echo when detuning the $\pi$-pulse from transition $f_{21}$. The $\pi$-pulse transfers the population from the excited state $e_1$ into the ground state $g_2$, partially suppressing the 2-level AFC echoes and enabling the spin-wave storage. When detuning greater than 100~kHz, the influence of control pulses diminish to a negligible level. Even though the $pi$-pulse is set to have a 1~MHz square bandwidth, its effective bandwidth is limited by the EDFA response and the limited Rabi frequency of the control pulse.

To directly observe the effect of inhomogeneous broadening in the hyperfine spin states, we create an AFC at $g_1$ with only antiholes by pumping transition $g_2 \rightarrow e_1$ (then decay to $g_1$), which is a more standard procedure in Eu$^{3+}$ or Pr$^{3+}$ (See Supp. Note S3).
Compared to the preparation method described in Fig.\ref{fig2}, we observe a significantly wider antihole linewidth of $\sim 130$~kHz (larger than a linewidth of $<$ 30 kHz of a direct hole burned).
We attribute this broadening to the involvement of the inhomogeneously broadened $g_1$–$g_2$ hyperfine transition during the population transfer process. In particular, the antihole formation relies on coherence between hyperfine levels with finite inhomogeneous width, leading to an effective broadening governed by the spin dephasing time $T_2^{*}$. As a result, the spectral features imprinted in $g_1$ inherit this additional broadening, in contrast to the narrower features obtained when the AFC is prepared directly within a single hyperfine manifold.

\begin{figure}[!h]
    \centering
    \includegraphics[width=0.8\linewidth]{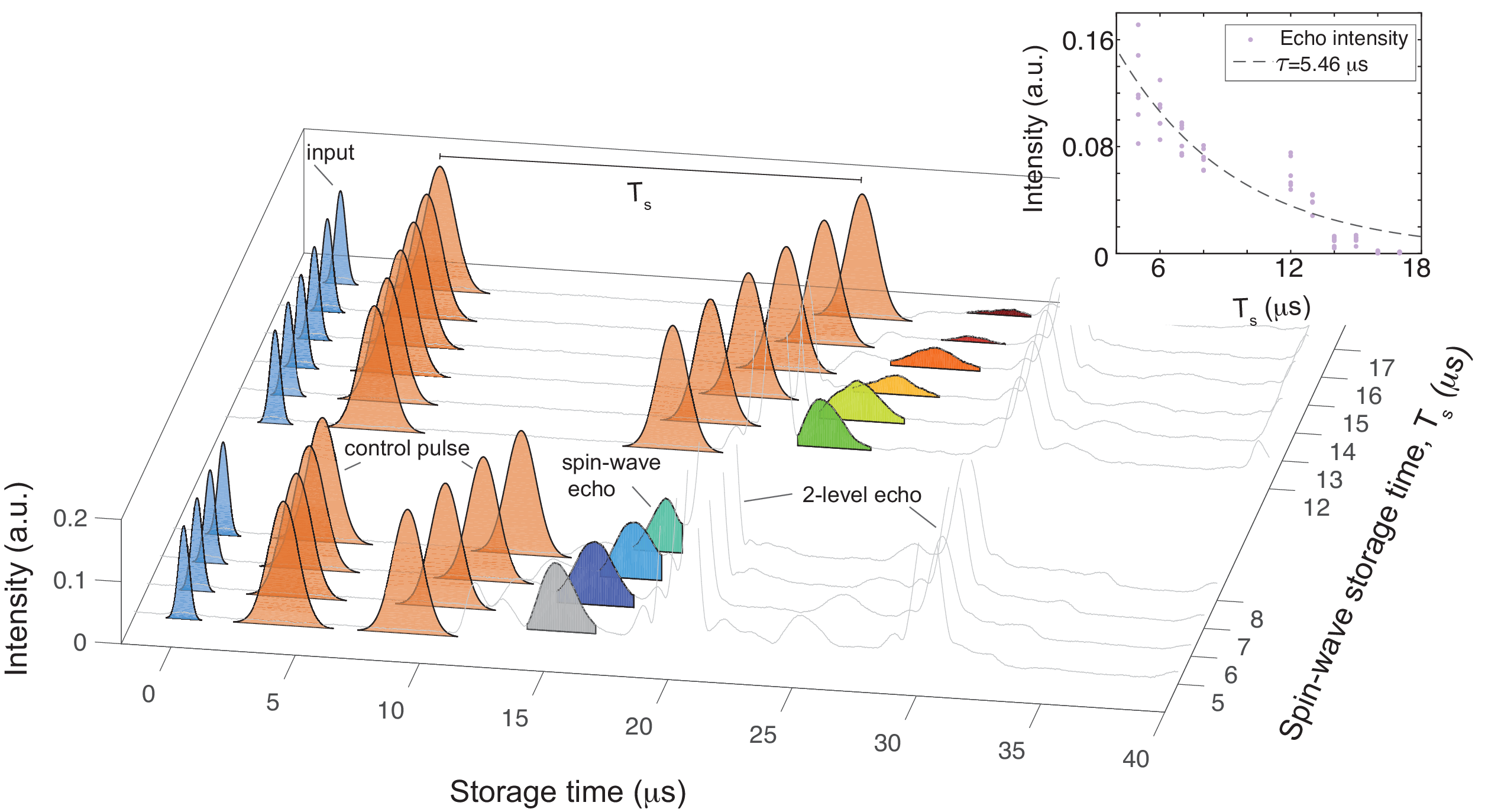}
    \caption{    
    Spin-wave echo intensity measured using lock-in detection as a function of storage time $T_s = 5,...,8~\mu s$ and $12,...,15~\mu s$.The signal decays with $1/e$ time $T_2^{*} = 5.46$~$\mu$s. The input and control pulses are shown schematically and do not represent the actual experimental pulses. Inset shows the raw memory output for different experiment runs taken at different spin-wave storage times, $T_s$ (varying the control pulse separation).}
  \label{fig4}
\end{figure}

Fig.~\ref{fig4} shows the spin-wave echo's intensity for different spin-wave storage times. The fitted decay time is about 5.5 $\mu$s.

\section{Discussion}
Quantum repeaters require on-demand storage of photonic qubits directly at telecom wavelengths to avoid lossy frequency conversion interfaces. While erbium-doped solids provide a uniquely compatible material platform, the absence of demonstrated spin-wave storage has prevented their use as true quantum memories. Our work closes this gap by establishing coherent transfer between optical and hyperfine excitations in Er-doped crystals and identifying the dominant mechanisms presently limiting memory performance.

A principal advantage of erbium-based memories is their direct operation in the telecom band, eliminating the need for quantum frequency conversion interfaces required by most alternative atomic memory platforms. In repeater architectures employing degenerate sources and non-telecom memories\cite{Albrecht:2014aa}, each memory interaction generally requires two conversion steps, yielding an overall efficiency $\eta_{\rm tot}
= \eta_{\rm mem}\,\eta_{\rm QFC}^{2},$ where $\eta_{\rm mem}$ is the intrinsic memory efficiency and $\eta_{\rm QFC}$ is the single-pass conversion efficiency. Even for optimistic noise-free $\eta_{\rm QFC}=0.8$, this introduces a $36\%$ loss per memory interaction. For a repeater chain of $M$ elementary links, this penalty scales exponentially as $\eta_{\rm chain}\propto (\eta_{\rm QFC}^{2})^M,$ yielding more than two orders of magnitude reduction for $M=10$. Native telecom memories avoid this multiplicative overhead, making erbium uniquely favorable for scalable long-distance quantum repeater architectures.

Moreover, the spin-wave protocol demonstrated here enables on-demand retrieval at
arbitrary times $T_s \gg 1/\Delta$ and has the potential to achieve
efficiencies approaching unity with cavity enhancement and improved
$\pi$-pulse transfer efficiency~\cite{Afzelius2009}. In this regard, we can compare the Er memory performance with photonic delay-line memories based on fiber loops\cite{Lee:2025aa} or free-space multi-pass cavities\cite{arnold2024all} as an alternative approach to temporary optical storage.  Such architectures can support exceptionally large optical bandwidths, potentially extending to the THz regime, however, information is stored in propagating photonic modes rather than stationary material excitations, causing the achievable storage time to scale directly with optical path length and accumulated round-trip loss, $\eta_{\rm delay}=\eta_{\rm rt}^{N_{\rm rt}},$ where $\eta_{\rm rt}$ is the round-trip efficiency and $N_{\rm rt}$ is the number of storage cycles. For example, considering a free-space delay cavity formed by mirrors with reflectivity $R=99.9\%$ separated by 10~m, the timing resolution given by the round-trip delay is $t_{\rm rt}=\frac{2L}{c}\approx 67~{\rm ns}.$
Achieving a storage time of order $100~\mu$s therefore requires $\sim1500$ round trips. Even neglecting diffraction and mode distortion, the accumulated transmission is limited to $\eta \approx R^{1500}\approx 22\%.$ In practice, spatial mode degradation, mirror aberrations, and switching imperfections further reduce the recoverable efficiency and fidelity after such large numbers of round trips.   

In contrast, atomic memories (with much smaller footprints) transfer the optical excitation to stationary hyperfine states, for which the storage time is fundamentally determined by the spin coherence time rather than optical propagation length. Retrieval is programmed directly by the control-pulse timing, enabling continuously tunable on-demand operation without repeated optical switching or increasing physical size with storage duration. This is a crucial property of memories enabling feed-forwarding for multiplexed processing in quantum networks\cite{Collins:2007aa}. Moreover, as we have shown below the Er atomic memory enables in-memory processing in both spectral and temporal domains\cite{li2025efficient}. 
 
The dominant limitation in the current experiment is $\pi$-pulse transfer efficiency,
which scales as $\eta_\pi \sim \sin^2(\Omega_R \tau_\pi / 2)$.
Increasing the amplifier output, reducing the beam waist, or using
composite pulse sequences~\cite{Levitt1986} will push $\eta_\pi$ toward
unity. Moreover, designing a longer 2-level echo storage enables application of longer control pulses, thus improving the transfer efficiency. 

Rančič \textit{et al.}~\cite{Rancic2018} showed that the hyperfine $T_2$
of $^{167}$Er:YSO can exceed 1~s at millikelvin temperatures and multi-tesla
fields using dynamical decoupling. The spin-wave storage has also been shown to be immune to additional noise thus enabling quantum-state storage\cite{gundougan2015solid, rakonjac2021entanglement}.

We estimate that moderate modification to the current experiment (currently under implementation) including 50\% increase in magnetic field and employing dynamical decoupling would increase the both the efficiency and storage time of the spin-wave storage time by an order of magnitude.
The current 2-level AFC efficiency of 9\% can be doubled by more efficiently initializing the spin population in a stronger magnetic field \cite{stuart2021initialization} or lower temperatures\cite{Wang:2025aa}. The transfer efficiency could be further enhanced by five-fold by increasing the pump power to 1~W.

At such conditions, the enhance $T_1$ time of the lower hyperfine levels
also favors complete optical pumping into $g_1$, improving the AFC contrast.  We note that all our measurements are performed in the linear storage regime, for which the Maxwell-Bloch evolution is independent of excitation photon number. 

Spin-wave storage in Er opens several avenues beyond AFC:
(i) \emph{DLCZ-type entanglement generation}~\cite{Duan2001} at telecom
wavelengths, with the read pulse now a resonant $\pi$ pulse\cite{Kutluer:2017aa}, rather than
a far-detuned Raman drive;
(ii) \emph{novel quantum memory protocols} where a combination of RF and optical pulses are used to implement a new class of light storage protocols~\cite{Moiseev:2025aa}; and
(iii) quantum transduction between microwave and optical fields mediated
by the spin degree of freedom~\cite{Williamson2014, Xie:2021aa}.
All of these protocols benefit directly from spin-wave storage demonstrated here.

In addition to spatial filtered employed here, further frequency filters are needed to the suppress control light to carry single-photon-level storage.  

Table \ref{tab:repeater} summarizes the important memory metrics of practical quantum repeaters demonstrated and reachable in the current platform. The current mode capacity estimated from the temporal modes that can be stored given the current delay-bandwidth product. The spectral mode capacity can be significantly improved by increasing the control pulse bandwidth.

\begin{table}[t]
\centering
\caption{Performance metrics relevant to telecom quantum repeater operation.}
\begin{tabular}{lccc}
\hline
Parameter & This work & Target regime & Improvement pathway \\
\hline
Storage time & $\sim 25~\mu$s & ms--s & Dynamical decoupling, ZEFOZ operation \\
Transfer bandwidth & $\sim 0.2-2$~MHz & 0.1 GHz & Pulse shaping, higher Rabi frequency \\
Transfer efficiency & $\sim 0.12$ & $>0.9$ & Optimal control pulses \\
Multimode capacity & $N\sim 10$ & $10^3$ & AFC bandwidth scaling \\
Telecom compatibility & Native & Required & Demonstrated \\
Cryogenic/B-field integration  & Compatible & Required & Demonstrated \\
\hline
\end{tabular}
\label{tab:repeater}
\end{table}

\section*{Conclusion}

In conclusion, we have demonstrated spin-wave AFC storage in Er$^{3+}$:YSO at telecom
wavelengths, enabled by strong control pulses
 that delivers sufficient Rabi frequency to drive coherent ground-state
hyperfine population transfer across the AFC bandwidth. 
We have characterized the spin-wave echo as a function of $\pi$-pulse power,
frequency, and storage time, and identified the ground-state inhomogeneous
broadening as a key parameter governing both the $\pi$-pulse bandwidth
requirement and the storage-time-dependent echo duration.
Our results show coherent optical-to-spin transfer in erbium as the key missing ingredient for realizing on-demand telecom quantum memories compatible with fiber-based quantum networking, and point toward storage times exceeding
1~s with higher magnetic fields and lower temperatures.


 \section*{Methods}
 {\bf Lock-in Detection.} The retrieved echo field is measured using heterodyne detection with a co-propagating local oscillator (LO) detuned by 200~MHz from the probe field. The detected photocurrent contains a beat note at the intermediate frequency $\Delta_{\rm LO}$,
\begin{equation}
I(t)\propto
{\rm Re}\!\left[E_{\rm LO}^*E_{\rm sig}(t)e^{-i\Delta_{\rm LO} t}\right],
\end{equation}
where $E_{\rm sig}(t)$ is the retrieved echo field.

The digitized signal is demodulated numerically into in-phase and quadrature components by mixing with both $\cos(\Delta_{\rm LO} t)$ and $\sin(\Delta_{\rm LO} t)$ reference signals, yielding
\begin{equation}
X(t)\propto {\rm Re}[E_{\rm sig}(t)],
\qquad
Y(t)\propto {\rm Im}[E_{\rm sig}(t)].
\end{equation}
The phase-insensitive echo amplitude is then reconstructed as
\begin{equation}
A(t)=\sqrt{X^2(t)+Y^2(t)}.
\end{equation}

This quadrature heterodyne measurement directly probes the coherent optical field rather than only its intensity and is insensitive to slow phase drifts between the signal and local oscillator. Observation of a stable heterodyne signal therefore confirms coherent retrieval of the stored optical excitation.
This measurement is formally equivalent to balanced optical heterodyne detection and directly probes the complex field coherence of the retrieved excitation rather than merely its intensity. 

{\bf Experimental details.}

 {\bf Spectral Preparation.}
 To establish a high-quality AFC within the $g_1-e_1-g_2\ \Lambda$-system, the preparation is executed as follows: 
(i). Concentrate the hyperfine population into the $\Lambda$-system by pumping $f_{i(i-1)}$ ($i=2-7$) and purify the target ion class by pumping $f_{11}$ and $f_{22}$, with a bandwidth of $\delta\omega=$7~MHz; 
(ii). Initialize those ions into $g_2$ by pumping $f_{11}$ ($\delta\omega=$7~MHz); 
(iii). Transfer ions back to $g_1$ by pumping $f_{21}$ ($\delta\omega=$2~MHz);
(iv). Generate AFC at $g_1$ by comb-pumping $f_{11}$ ($\delta\omega=$2~MHz), meanwhile cleaning $g_2$ by pumping $g_2 \leftrightarrow e_2$ ($\delta\omega=$7~MHz).
We optimize the number of pumping cycles to improve the population initialization and comb contrast, yielding a well-defined $\Lambda$ system for subsequent spin-wave storage.

 {\bf Control Pulse Power Requirement.}
 In the two-level AFC protocol, an input pulse absorbed at time $t=0$ re-emits
an echo  at $t = 1/\Delta$ ($\Delta$: tooth spacing).
A control ($\pi$) pulse applied at $t = \tau$ (before re-emission) maps the optical coherence onto a
spin coherence (spin wave), in principle, silencing the two-level echo.
A second $\pi$ pulse at $t = \tau + T_s$ maps it back, yielding a
spin-wave echo at $t = 1/\Delta + T_s$, where $T_s$ is the controllable storage time~\cite{Afzelius2009}.

For the spin-wave echo to be observable, the $\pi$ pulse must achieve
a transfer efficiency $\eta_\pi$ close to unity over the entire AFC
bandwidth $\Gamma_{\rm AFC}$.
The time-bandwidth product of the $\pi$ pulse is constrained by
\begin{equation}
  \Omega_R \, \tau_\pi \approx \pi, \qquad
  \Omega_R \gg 2\pi\,\Gamma_{\rm AFC},
  \label{eq:pi_bandwidth}
\end{equation}
where $\Omega_R = \mu_{ge} \mathcal{E}_\pi / \hbar$ is the Rabi frequency
and $\mathcal{E}_\pi$ is the $\pi$-pulse field amplitude.
This imposes a minimum optical power needed to effectively transfer the excitations to/from the ground-state spin.


Using intensity $I = \frac{1}{2} c \varepsilon_0 (\hbar \Omega_R / \mu)^2$, for Er$^{3+}$ ions in YSO, with a typical optical dipole moment $\mu \sim 10^{-32}$~C$\cdot$m and an AFC bandwidth in the MHz range, we estimate intensities on the order of $10^8$~W/m$^2$. For a Gaussian beam with a waist of $100~\mu$m, the required optical power is on the order of $\sim$1~W .

\bibliography{main}

\section*{acknowledgments}
Authors like to thank Yisheng Lei and Kim Fook Lee for early discussions and brainstorming. We like to thank funding from U.S. Department of Energy, Office of Science, Office of Advanced Scientific Computing Research, through the Quantum Internet to Accelerate Scientific Discovery Program under Field Work Proposal No. 3ERKJ381.

\subsection*{Contributions}
Z.L. has implemented the experiment, collected and analyzed data. Both authors interpreted the data and prepared the manuscript.


\clearpage
\newpage
\appendix
\setcounter{equation}{0}
\setcounter{figure}{0}
\setcounter{table}{0}
\renewcommand{\theequation}{A\arabic{equation}}
\renewcommand{\thefigure}{A\arabic{figure}}
\renewcommand{\thetable}{A\arabic{table}}

\section*{Supplementary Material: On-Demand Coherent Mapping of Telecom Optical States onto Erbium Hyperfine Spins}

\section*{S1. Hole-burning spectra at multiple frequencies}

\begin{figure}[h!]
    \centering
    \includegraphics[width=0.8\linewidth]{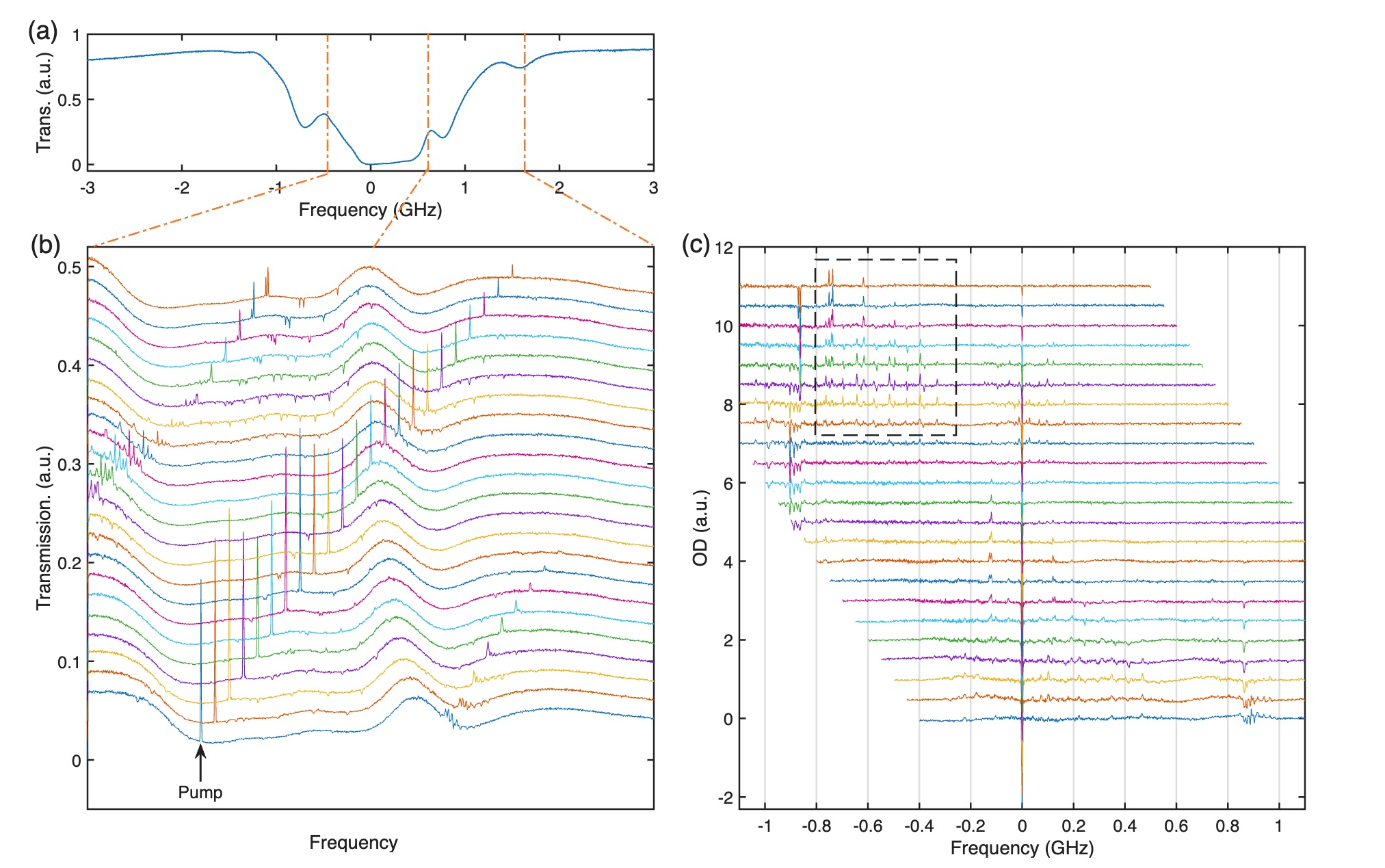}
    \caption{(a) Erbium transmission spectrum centered at 1536.31~nm. (b) Transmission spectra showing hole and antiholes created by pump at different frequencies (shifted vertically for better visibility). 
    (c) spectra in (b) transferred to optical depth and centered around the pump frequency. The dashed box shows the main antiholes reflecting the hyperfine level structures.
    }
    \label{fig:holespec}
\end{figure}

Figures~\ref{fig:holespec} shows hole-burning spectra recorded at different pump frequencies. 
The continuously varying peaks on each curve in Fig.~\ref{fig:holespec}(b) are the hole burning frequencies. 
Fig.~\ref{fig:holespec}(c) shows the corresponding optical depth spectra and aligned with the burning frequency.
The side-holes at -900 to -870~MHz (away from the main hole) indicate the excited state energy gaps.
The antiholes at $\sim -750$ MHz reflect the ground state energy gaps. 
The antiholes between -700 MHz and -300 MHz are those ions for which pump is at $g_{i+1}\rightarrow e_{i}$ ($f_{i+1,i}$, $i=1,...,7$) and probe at $g_1 \rightarrow e_1$ ($f_{11}$). 
The antihole structure reveals that at the edge of the inhomogeneous broadening (higher pumping frequency, top curve of the Fig.~\ref{fig:holespec}(b)), only upper ground state ($g_1$, $g_2$) of the ions could be excited by the laser. As pumping frequency closing to the center, more ground state are involved and more antihole appears.
This enables us to distinguish the energy-level structure of the ions, which is given in Table~\ref{tab}.
With those information, we can concentrate the population in 7 out of 8 hyperfine levels into the our $\Lambda-$system.
The transmission spectra in (b) is through a $\sim$0.5 GHz linewidth etalon, as such it looks different from the spectrum in (a).

\begin{table}[h!]
    \centering
    \begin{tabular}{|c|c|}
\hline
Two levels or transition &   Frequency difference (MHz) \\
\hline
 $g_1-g_2$ &  751.1 \\
 $g_2-g_3$ & 736.8 \\
 $g_3-g_4$ & 741.3 \\
 $g_4-g_5$ & 763.3 \\
 $e_1-e_2$ & 870.7 \\
 $e_2-e_3$ & 863.0 \\
 $e_3-e_4$ & 862.0 \\
 $f_{65}-f_{11}$ & 330.4 \\
 $f_{76}-f_{11}$ & 302.0 \\
\hline
    \end{tabular}
    \label{tab}
    \caption{The energy gap or transition difference of the Er$^{3+}$ identified in the experiment.}
\end{table}

\newpage
\section*{S2. Anti-hole lifetime versus spectral position}

\begin{figure}[h!]
    \centering
    \includegraphics[width=0.5\linewidth]{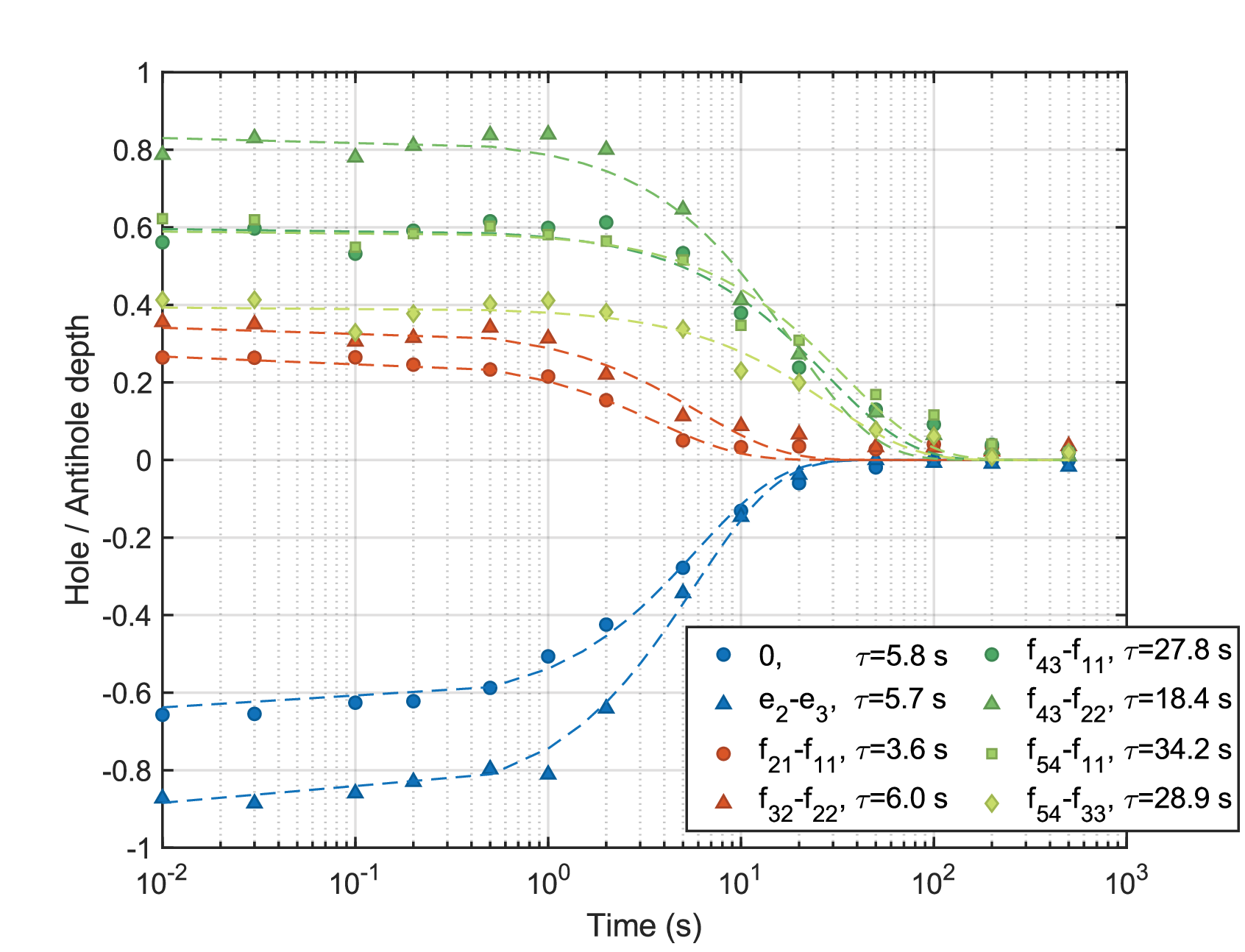}
    \caption{Hole depth decay curve of relatively stronger antiholes (depth$>0$) and holes (depth$<0$) are plotted. Exponential fittings shown as dashed lines give different hole lifetimes. }
    \label{fig:holelifetime}
\end{figure}

We have observed that the holes and antiholes corresponding to different transitions in a hole-burning experiment have different lifetimes. 
Figure~\ref{fig:holelifetime} plots the lifetime of different antiholes (depth$>0$) and holes (depth$<0$) with the pumping frequency corresponding to the 4th curve from the top in Fig.~\ref{fig:holespec}(c). 
The antiholes of pumping $f_{(i+1)i}$ and probing $f_{ii}$ (denote as $f_{(i+1)i} - f_{ii}$) and holes have a typical lifetime of 3~s, in agreement with the hyperfine relaxation time $T_1$. The decay  is dominated by the phonon coupling as explained in our last paper \cite{li2025efficient}.
The antiholes of $f_{(i+1)i} - f_{jj}$ ($i>j$) have a much longer lifetime (20 - 30 s). Because the population are emptied at $g_{i+1}$ and accumulated at $g_i$ to $g_j$, the the antihole $g_j$ level don't have an adjacent empty level and thus decay slower. This is also reflected in the longer lifetime of $f_{(i+1)i}-f_{11}$ comparing to $f_{(i+1)i}-f_{22}$ because of the $g_1$ more far away from the empty level $g_{i+1}$.
Additionally, we observe that the antihole lifetime is transition dependent instead of being frequency dependent. This is confirmed by making the antihole $f_{32}-f_{11}$ and antihole $f_{21}-f_{11}$ located on the same spectrum in two separate hole burning experiments with different pump frequencies. This gives a corresponding hole lifetime of 25~s and 3~s, respectively. 
The variation is accounted for when choosing the AFC preparation frequency to maximize the useful spin-wave storage window.
 
\section*{S3. Hyperfine inhomogeneous broadening}

\begin{figure}[h]
    \centering
    \includegraphics[width=0.5\linewidth]{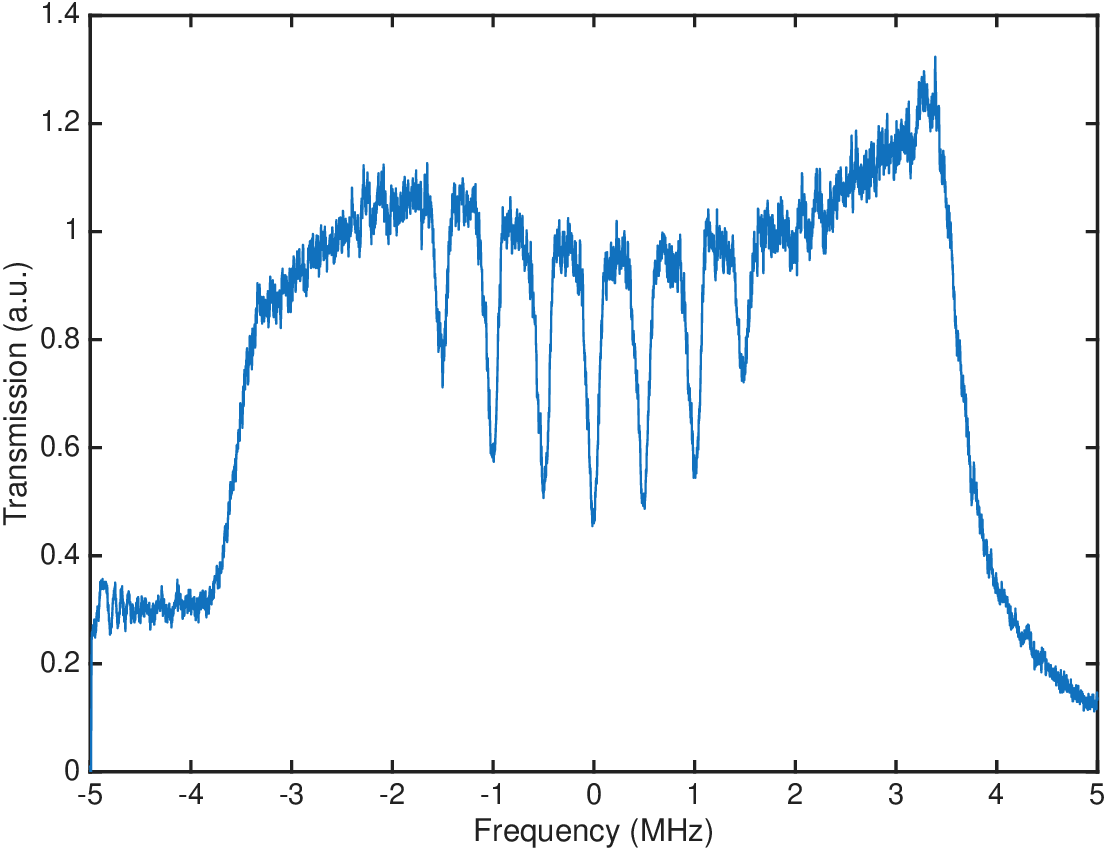}
    \caption{AFC composed of antiholes (dips in the figure). The minimum linewidth of antihole is limited to 130~kHz due to the hyperfine inhomogeneous broadening.}
    \label{fig:spin_inh}
\end{figure}

In this work, we construct the AFC in $g_1$ by burning holes at $g_1$ instead of cleaning $g_1$ and then pumping back through $f_{21}$. The later is more standard and widely used in Eu$^{3+}$ and Pr$^{3+}$ spin-wave storage.
This is because the hyperfine inhomogeneous broadening is measured to be 130~kHz, which is given by the minimum linewidth of antihole, as shown in Fig.~\ref{fig:spin_inh}.
The antihole FWHM of 130 kHz limits the 2-level AFC storage time to $1/(2*130\ \mathrm{kHz}) = 3.8\ \mu$s, which is too short to demonstrate the spin-wave storage. In contrast, the minimum \textit{hole} linewidth of $< 30$~kHz is more feasible for a 10~us AFC for the spin-wave memory.

\end{document}